\documentclass[prd,superscriptaddress,amsmath,twocolumn,nofootinbib]{revtex4-2}

\pdfoutput=1
\usepackage[utf8]{inputenc}
\usepackage{microtype}
\usepackage{times}
\usepackage{graphicx}
\usepackage{amssymb,mathtools}
\usepackage{enumitem}
\usepackage{xcolor}
\usepackage{floatrow}
\usepackage{nicefrac}
\usepackage{tensor}

\def\sg{\textsl{g}}
\newcommand{\RNum}[1]{\uppercase\expandafter{\romannumeral #1\relax}}

\makeatletter
\newcommand*{\defeq}{\mathrel{\rlap{%
	\raisebox{0.3ex}{$\m@th\cdot$}}%
	\raisebox{-0.3ex}{$\m@th\cdot$}}%
	=}
\newcommand*{\eqdef}{=\mathrel{\rlap{%
	\raisebox{0.3ex}{$\m@th\cdot$}}%
	\raisebox{-0.3ex}{$\m@th\cdot$}}%
	}
\makeatother

\usepackage{etoolbox}
\patchcmd{\section}
{\centering}{\raggedright}{}{}
\patchcmd{\subsection}
{\centering}{\raggedright}{}{}

\usepackage{scalerel}
\usepackage{tikz}
\usetikzlibrary{svg.path}
\definecolor{orcidlogocol}{HTML}{A6CE39}
\tikzset{
	orcidlogo/.pic={
		\fill[orcidlogocol] svg{M256,128c0,70.7-57.3,128-128,128C57.3,256,0,198.7,0,128C0,57.3,57.3,0,128,0C198.7,0,256,57.3,256,128z};
		\fill[white] svg{M86.3,186.2H70.9V79.1h15.4v48.4V186.2z}
		svg{M108.9,79.1h41.6c39.6,0,57,28.3,57,53.6c0,27.5-21.5,53.6-56.8,53.6h-41.8V79.1z M124.3,172.4h24.5c34.9,0,42.9-26.5,42.9-39.7c0-21.5-13.7-39.7-43.7-39.7h-23.7V172.4z}
		svg{M88.7,56.8c0,5.5-4.5,10.1-10.1,10.1c-5.6,0-10.1-4.6-10.1-10.1c0-5.6,4.5-10.1,10.1-10.1C84.2,46.7,88.7,51.3,88.7,56.8z};
	}
}
\newcommand\orcidlink[1]{\href{https://orcid.org/#1}{\mbox{\scalerel*{
				\begin{tikzpicture}[yscale=-1,transform shape]
					\pic{orcidlogo};
				\end{tikzpicture}
			}{X}}}}

\usepackage{url,hyperref}
\hypersetup{colorlinks,linkcolor={blue!55!black},citecolor={red!45!black},urlcolor={blue!45!black},breaklinks=true}

\begin{document}

\title{Nomen non est omen: \\ Why it is too soon to identify ultra-compact objects as black holes}

\author{Sebastian Murk\orcidlink{0000-0001-7296-0420}}
\email{sebastian.murk@oist.jp}
\affiliation{Okinawa Institute of Science and Technology, 1919-1 Tancha, Onna-son, Okinawa 904-0495, Japan}
\affiliation{School of Mathematical and Physical Sciences, Macquarie University, Sydney, New South Wales 2109, Australia}
\affiliation{Sydney Quantum Academy, Sydney, New South Wales 2006, Australia}

\begin{abstract}
	Black holes play a pivotal role in the foundations of physics, but there is an alarming discrepancy between what is considered to be a black hole in observational astronomy and theoretical studies. Despite claims to the contrary, we argue that identifying the observed astrophysical black hole candidates as genuine black holes is not justified based on the currently available observational data, and elaborate on the necessary evidence required to support such a remarkable claim. In addition, we investigate whether the predictions of semiclassical gravity are equally compatible with competing theoretical models, and find that semiclassical arguments favor horizonless configurations.
\end{abstract}

\maketitle

\section{Introduction} \label{sec:introduction}
Black holes are arguably the most celebrated prediction of general relativity. Initially regarded merely as mathematical curiosities, steady advances in the precision of astronomical observations have gradually shifted our perception of black holes from purely mathematical objects to potentially real physical entities. The last few decades in particular have produced strong evidence for the existence of dark massive compact objects commonly referred to as ``astrophysical black hole candidates'' (or simply ``astrophysical black holes''). Prominent examples include the tracking of stellar orbits \cite{g:08,g:09}, detection of gravitational waves by laser interferometry \cite{LIGO:Virgo:16,LIGO:Virgo:KAGRA:21}, and imaging of the photosphere/shadow by very-long-baseline-interferometry \cite{EHT:19,EHT:22}. Nonetheless, the true physical nature of these objects remains unknown \cite{v:08,h:14,f:14,bmt:17,cp:19}: contemporary models describe them either as horizonless ultra-compact objects (UCOs) or genuine black holes with some type of horizon.

If black holes exist, we must come to terms with the consequences of their existence, such as the nontrivial causal structures they inevitably introduce into the spacetime at large \cite{p:65,sg:15,l:22}. Somewhat surprisingly, despite the profound impact on our understanding of fundamental physics, the not-so-subtle differences between black holes and horizonless UCOs are often ignored in contemporary scientific endeavors. It is frequently asserted (see Refs.~\citenum{EHT:22} and \citenum{EHT:press:22} for recent claims) that the presently available results provide ``direct evidence'' that the observed astrophysical black hole candidates are indeed black holes as opposed to alternative descriptions of dark massive compact objects. The aim of this short report is to outline the physical implications of these claims.

The remainder of this article is organized as follows: in Sec.~\ref{sec:BHzoo}, we review the classification scheme for different types of UCOs, highlight their differences, and clarify what empirical evidence is required to identify the observed astrophysical black holes candidates as black holes. In Sec.~\ref{sec:SCconsiderations}, we investigate whether the predictions of semiclassical gravity are equally compatible with competing theoretical models. In Sec.~\ref{sec:soa}, we briefly summarize the observational state of affairs and examine whether any of the characteristic properties of distinct theoretical models can be identified with the currently available astrophysical data. Lastly, we discuss the consequences of our findings and comment on promising avenues for future observations and theoretical considerations (Sec.~\ref{sec:conclusions}).

\section{Navigating the ultra-compact object zoo} \label{sec:BHzoo}

\subsection{Black holes and their horizons}
To assess the validity of the claim that the observed objects are black holes, we must have a clear understanding of what makes a black hole a black hole, and, in particular, what distinguishes a black hole from alternative descriptions of dark massive compact objects such as horizonless UCOs. While there is no unanimously agreed upon definition of a black hole \cite{c:19}, trapping of light is a commonly accepted feature that naturally emerges in all conventionally relevant contexts. The notion of a trapped spacetime domain can be expressed mathematically using the concept of a closed trapped surface pioneered by Sir Roger Penrose \cite{p:65,p:68}. The outermost boundary of a trapped spacetime region, from which not even light can escape, is popularly referred to as its horizon. A well-known example is the event horizon; it is the boundary of the black holes predicted by general relativity (so-called mathematical black holes, abbrev.\ MBHs). Schwarzschild and Kerr black holes are typical examples, and astrophysical observations are typically modeled based on the Schwarzschild/Kerr paradigm.

However, it is worth pointing out that the event horizon, while mathematically convenient, is a highly idealized global definition of the boundary of a black hole. As such, it requires knowing the entirety of spacetime, either from initial data or by direct construction \cite{l:21}. As pointed out in Ref.~\citenum{v:14}, this implies that ``one needs to know the entire history of the Universe, all the way into the infinite future, and all the way down to any spacelike singularity, to decide whether or not an event horizon exists''. Consequently, it is impossible for us as quasilocal observers (who do not have access to global topological information) to determine the existence, presence, or absence of an event horizon: they are not physically observable (not even in principle) and thus of severely limited suitability for practical purposes such as observational astronomy \cite{cp:19,akl:02,v:14}. In contrast, there are several well-defined quasilocal horizon notions (e.g.\ apparent horizon, dynamical horizon, trapping horizon) whose presence or absence can (at least in principle) be determined observationally \cite{ak:04,v:14,s:rev:11,f:14,f:book:15}. Formal mathematical definitions of various horizons are provided in Appendix A.1 of Ref.~\citenum{mmt:rev:22}. Following the terminology of Ref.~\citenum{f:14}, we refer to trapped spacetime regions that are bounded by physically observable horizons as physical black holes (PBHs). Singularity-free PBH solutions are referred to as regular black holes (RBHs). However, note that their dynamical evolution is severely constrained by self-consistency requirements, and their stability requires further investigation, as detailed in Refs.~\citenum{pi:89,pi:90,o:91,ha:10,dmt:22,cdlv:22a}.

Since the presence of a horizon is the principal characteristic that distinguishes genuine black holes from horizonless UCOs that closely mimic their properties, this is the necessary evidence needed to support any claim that the observed astrophysical black hole candidates are indeed black holes. According to Earman’s principle ``no effect can be counted as a genuine physical effect if it disappears when the idealizations are removed'' \cite{e:04}. Consequently, in order for a horizon to be considered a genuine physical object rather than merely a useful mathematical tool, it must form in finite time of a distant observer, and there should be some at least in principle observable consequences of its formation \cite{mt:21cp,dmt:22}. In fact, there are various proposals for detecting the presence or absence of a horizon based on potentially observable features, including but not limited to differences in the quasinormal mode spectrum (most notably gravitational wave echoes \cite{cp:17,h:20,lac:21}) and properties of accretion flows \cite{af:13,v:21,cdlv:22b}.

\subsection{Classification of ultra-compact objects}
\begin{figure*}
	\floatbox[{\capbeside\thisfloatsetup{capbesideposition={right,center},capbesidewidth=68.7mm}}]{figure}[\FBwidth]
	{\hspace*{-4.6mm} \caption{Classification of dark compact objects by their compactness $r_\sg/r_0=1/(1+\epsilon)$. Genuine black holes --- whether mathematical (MBHs), physical (PBHs), or regular (RBHs) --- correspond to $\epsilon=0$. The trapped spacetime domains of PBH and MBH solutions overlap, but neither is contained within the other. Planck-scale corrections correspond to $\epsilon \lesssim 10^{-40}$, and clean photosphere objects with $\epsilon \lesssim 0.0165$ are expected to mirror the early-time dynamics of black holes (for details, see Refs.~\citenum{cp:17,cp:19,bcns:19}). Models of horizonless UCOs with $\epsilon < 1/8$, such as exotic compact objects, violate at least one assumption of the Buchdahl theorem \cite{b:59}. Figure adapted from Fig.~1 of Ref.~\citenum{mmt:rev:22}.} 
		\label{fig:UCOclassification}}
	{\includegraphics[scale=0.77]{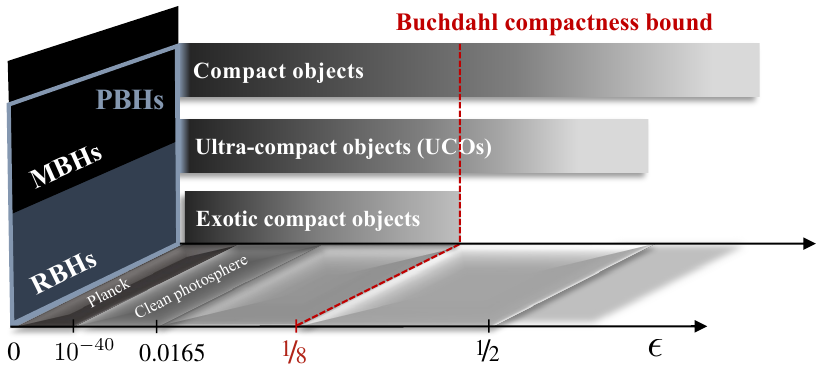}}
\end{figure*}

A rudimentary classification scheme of dark compact objects is presented in Fig.~\ref{fig:UCOclassification}. To quantify deviations from black hole spacetimes, it is customary to introduce a so-called closeness parameter $\epsilon$ \cite{cp:19}. Many definitions are possible in principle, but in spherical symmetry Birkhoff's theorem \cite{j:21,b:23} provides a natural choice via $r_0/r_\sg \eqdef 1 + \epsilon$, where $r_\sg$ denotes the Schwarzschild radius of the object, and $r_0$ the radius of its effective surface. Black holes --- irrespective of what type --- correspond to $\epsilon = 0$ (cf.\ Fig.~\ref{fig:UCOclassification}). Models of horizonless UCOs with closeness $\epsilon < 1/8$, such as exotic compact objects, violate at least one assumption of the Buchdahl theorem \cite{b:59} that bounds the compactness of self-gravitating objects by $r_\sg/r_0 < 8/9$.

\subsection{Gravitational collapse scenarios} \label{subsec:grav-collapse-scenarios}
From the point of view of a distant observer, gravitational collapse beyond the density of a neutron star can have three possible outcomes \cite{mmt:rev:22}:
\begin{enumerate}[label=\Roman*.]
	\item Formation of a transient object: the closeness parameter reaches a minimal value $\epsilon_\text{min} > 0$, either at some finite time $t_\text{min}$ or asymptotically as $t \to \infty$.
	\item Perpetual ongoing collapse ($\epsilon \to 0$ as $t \to \infty$): the horizon exists only as an asymptotic ($t \to \infty$) concept. At times $t < \infty$, objects of this type are \textit{de facto} horizonless.
	\item Formation of an apparent horizon in finite asymptotic time $t_\mathrm{S} < \infty$, i.e.\ in finite time according to the clock of a distant observer. 
\end{enumerate}
Since our measurements occur at $t_\mathrm{S} < \infty$, scenario \RNum{3} is the only one compatible with the claimed observation of black holes. We thus focus on this outcome in what follows. Note that known classical models of gravitational collapse describe formation of the event horizon as an asymptotic event for a distant observer, and thus correspond to scenario \RNum{2}. Indeed, only models with matter that violates the null energy condition (NEC) \cite{he:book:73,bhl:18,bmmt:19} (see Sec.~\ref{subsec:horizon-formation}) may describe the formation of trapped regions in finite time of a distant observer (scenario \RNum{3}), indicating that quantum effects are needed at least at the level captured by semiclassical gravity.

\section{Semiclassical considerations} \label{sec:SCconsiderations}
Semiclassical gravity \cite{hv:book:20} is often considered to be the best-tested framework of gravitational physics we currently have available. One of its most impressive achievements is the prediction that black holes evaporate through the emission of Hawking (i.e.\ completely thermal) radiation \cite{h:74,h:75}, thereby giving rise to black hole thermodynamics \cite{w:01,p:05} and the infamous information loss problem \cite{m:rev:17,uw:rev:17}.

\subsection{Semiclassical Einstein equations}
Within this framework, the dynamics is described by the semiclassical Einstein equations
\begin{align}
	\tensor{R}{_\mu_\nu} - \frac{1}{2} R \tensor{\sg}{_\mu_\nu} = 8 \pi \langle \tensor{\hat{T}}{_\mu_\nu} \rangle_\psi ,
	\label{eq:semiclassicalEE}
\end{align}
where the use of Planck units $G=\hbar=c=1$ is implied, the cosmological constant term was omitted, and $\langle \tensor{\hat{T}}{_\mu_\nu} \rangle_\psi \defeq \langle \psi \vert \tensor{\hat{T}}{_\mu_\nu} \vert \psi \rangle$ denotes the expectation value of the renormalized energy-momentum tensor, a quantum field theory operator that describes both the collapsing matter and the quantum field excitations it produces. This joint treatment of the entire matter content is the hallmark feature of the self-consistent semiclassical approach \cite{mmt:rev:22}. Since spacetime geometry described by the LHS of Eq.~\eqref{eq:semiclassicalEE} is still classical, it is implicitly assumed that classical notions such as the metric, horizons, and trajectories continue to be physically meaningful. However, no \textit{a priori} assumptions are made about the matter content of the theory (including the status of energy conditions), the quantum state $\psi$, or the asymptotic structure of spacetime. 

In the following subsections Sec.~\ref{subsec:horizon-formation} and Sec.~\ref{subsec:accretion}, we analyze the physical consequences of having gravitational collapse according to scenario \RNum{3} of Sec.~\ref{subsec:grav-collapse-scenarios} realized in the semiclassical theory. However, note that the argumentation presented therein may not pertain to primordial black holes as the precise details of their formation are not known.

\subsection{Horizon formation} \label{subsec:horizon-formation}
If the semiclassical theory is valid, the formation of trapped regions in finite time of a distant observer requires a violation of the NEC \cite{he:book:73,bhl:18,bmmt:19} near the outer apparent horizon located at the Schwarzschild radius $r_\sg$ \cite{fethm:17}. The NEC is the weakest of all energy conditions: it postulates that $\tensor{T}{_\mu_\nu} \ell^\mu \ell^\nu \geqslant 0$, that is the contraction of the energy-momentum tensor with any future-directed null vector $\ell^\mu$ is non-negative. Its violation implies the existence of macroscopic regions of negative energy density, i.e.\ “matter” that behaves unlike anything we have observed thus far. In other words, the formation of an observable horizon in finite time of a distant observer requires exotic matter. Despite the fact that this has been known for roughly 50 years (at least since Ref.~\citenum{he:book:73}), this result is often either ignored or overlooked in contemporary literature. Since the NEC is the weakest of all energy conditions, its violation implies that all other energy conditions (strong, weak, dominant) are violated as well. Furthermore, since the proofs of the singularity theorems \cite{s:98} are premised on the NEC (or one of the stronger energy conditions) being satisfied, there is no guarantee that they will continue to hold. However, recall that while singularity-free RBHs are a conceivable outcome of gravitational collapse in principle, the stability of dynamical models is subject of contemporary debate \cite{pi:89,pi:90,o:91,ha:10,dmt:22,cdlv:22a}.

For macroscopic black holes ($r_\sg \gg 1$) such as the observed candidate objects, thin shell models \cite{p:book:04} of gravitational collapse are expected to provide a good approximation for the timescale of horizon formation and evaporation \cite{bmt:19}. However, estimates of the energy and timescale required for the formation of an apparent horizon in finite time of a distant observer (scenario \RNum{3} of Sec.~\ref{subsec:grav-collapse-scenarios}) are incompatible with the identification of the observed objects as black holes, i.e.\ even if horizon formation is possible in principle, it is much too early \cite{mt:21}: for a solar mass black hole this time is about $t_{\odot} \sim 10^{67}$yr, and for the compact objects Sgr A* and M87* with masses $M_{\text{Sgr A*}} \approx 4 \cdot 10^6 M_\odot$ \cite{EHT:22} and $M_{\text{M87*}} \approx 6.5 \cdot 10^{9} M_\odot$ \cite{EHT:19}, one finds $t_{\text{Sgr A*}} \sim 10^{86}$yr and $t_{\text{M87*}} \sim 10^{96}$yr, respectively. All of these estimates exceed the age of our universe $t_{\text{u}} \sim 13.8 \cdot 10^9$yr considerably.

\subsection{Accretion of matter: Crossing the horizon} \label{subsec:accretion}
In spherical symmetry, accretion of matter in the sense that matter crosses the horizon and increases the mass of the PBH is no longer possible after the horizon has been formed, i.e.\ PBHs --- if they exist --- can only evaporate. This is due to the presence of firewalls that violate quantum energy inequalities \cite{f:book:17,ks:20} that bound the extent of the NEC violation. A detailed derivation is provided in Refs.~\citenum{mmt:rev:22} and \citenum{dmt:22}. 

In contrast, LIGO/Virgo gravitational-wave interferometers have recorded merger events in which the resulting object's mass exceeds each of its progenitor's masses, indicating that some mass must have been acquired, and thus some horizon (if the observed black hole candidates possess them as claimed) must have been crossed. Therefore, avoiding contradictions with observational data requires some physical effect that does not exist in spherical symmetry, but is present in axially symmetric spacetimes used to describe the emission of gravitational waves by rotating Kerr--Vaidya black holes \cite{dt:20} and can explain the claimed crossing of the horizon.

Similarly, EHT results demonstrate that both Sgr A* and M87* accrete matter in the vicinity of their would-be horizons \cite{EHT:19,EHT:22}. Since the currently available observational data is insufficient to infer the presence of a horizon, the crossing of matter cannot be inferred either. While it is conceivable in principle that matter accretes in the vicinity of the horizon (if it exists) without crossing it, we are not aware of any physical mechanism that would enable continued accretion of matter very close to the horizon while at the same time prohibiting the flow of matter across it.

\section{State of affairs} \label{sec:soa}
From an observational point of view, the current state of affairs may be summarized as follows:\footnote{For a more detailed account, the interested reader is referred to Refs.~\citenum{cp:19,cp:17,lac:21,v:21,cdlv:22b,ccy:22}.}
\begin{enumerate}
	\item Both models used to describe astrophysical black hole candidates, that is black holes and horizonless UCOs, are compatible with all currently available astrophysical and cosmological data.
	\item The sensitivity of present-day astronomical observations, including the most recent observations by the EHT and LIGO/Virgo Collaborations, is insufficient to infer the existence (or non-existence) of a horizon.
\end{enumerate}
This suffices to refute any claim that the presently available results constitute ``direct evidence'' \cite{EHT:22,EHT:press:22} that would justify identifying the observed astrophysical black hole candidates as genuine black holes (irrespective of what type, cf.\ Fig.~\ref{fig:UCOclassification}). Nonetheless, it may prove fruitful to carefully reflect on the following additional considerations.

According to the principle of parsimony (also known as ``Ockham's razor''), if two (or more) models explain observations with the same level of accuracy, the simpler (simplest) model is to be preferred. A popular argument against horizonless UCOs is that their compactness is limited by the Buchdahl theorem \cite{b:59}, i.e.\ the formation of extremely compact horizonless objects (those with $\epsilon < 1/8$, see Fig.~\ref{fig:UCOclassification}) requires violating at least one of its assumptions. However, as outlined in Sec.~\ref{subsec:horizon-formation}, the formation of a horizon in finite time of a distant observer inevitably requires negative-energy-density exotic matter. Therefore, PBHs are at least as (if not more) exotic as horizonless UCOs. In fact, given that 
\begin{enumerate}[label=\alph*.]
	\item violating the NEC is not a mathematical prerequisite for the formation of extremely compact horizonless UCOs: exotic matter can be avoided provided that at least one assumption of the Buchdahl theorem \cite{b:59} (e.g.\ spherical symmetry, matter being described by a single perfect fluid, isotropy of the fluid, an energy density profile that decreases as one moves outwards ($d\rho(r)/dr<0$), etc.) is violated. As demonstrated in Refs.~\citenum{abcg:21a} and \citenum{abcg:21b}, semiclassical corrections ineluctably lead to violations of several of these assumptions, thus providing multiple ways of surpassing the Buchdahl compactness bound without invoking new physics or the presence of exotic matter;
	\item problems arising from the separation of spacetime innate to black holes (such as the presence of singularities where the theory breaks down or the apparent lack of unitarity in the aforementioned information loss problem) are naturally avoided in models of horizonless UCOs;
\end{enumerate}
one may be inclined to argue that horizonless UCO models are simpler and/or more compatible with the predictions of quantum field theory in curved spacetimes, and thus to be preferred.

\section{Conclusions} \label{sec:conclusions}
Massive dark UCOs are one of the most promising areas of gravitational physics research. Due to their compact nature and strong gravitational fields, they maximally highlight differences in the predictions of general relativity and alternative theories of gravity, making them excellent laboratories to probe fundamental physics. 

Despite claims that the observed astrophysical black hole candidates are black holes, there is currently no evidence that any one of them possesses a horizon. Both genuine black holes and horizonless UCOs are compatible with observational data, but the observation of PBHs (those bounded by quasilocal and in principle observable horizons) requires never-before-seen exotic matter and appears to be incompatible with the predictions of semiclassical gravity (see Sec.~\ref{subsec:horizon-formation} and Sec.~\ref{subsec:accretion}). In this sense, the detection of observational signatures indicative of a horizon would provide indubitable evidence for the breakdown of semiclassical physics at the horizon scale. 

In any case, the act of insisting on new physics beyond what has been observed should not be taken lightly. It is typically reserved as a means to overcome or (attempt to) alleviate unresolved issues, such as to reconcile discrepancies with observations or resolve mathematical inconsistencies. Candidate theories of quantum gravity for instance are generally expected to cure the singularities predicted by general relativity. However, here, insisting on the identification of the observed astrophysical black hole candidates with PBHs (i.e.\ insisting on exotic matter or accepting the breakdown of semiclassical gravity) has the opposite effect: instead of resolving outstanding issues, the presence of a horizon introduces problems (see item b. in Sec.~\ref{sec:soa}) that are absent in horizonless UCO models (which, to reiterate, are also compatible with observational data and --- despite the commonly used terminology ``exotic compact objects" to refer to UCOs with compactness $\epsilon < 1/8$ --- do not require exotic matter to be compatible with semiclassical gravity). Even if the necessary violation of the NEC occurs in nature, the process of horizon formation appears to be too slow to justify identifying the presently observed objects as black holes (see Sec.~\ref{subsec:horizon-formation}).

Increases in the sensitivity of future observations combined with improvements in the precision of predictions of competing theoretical models are expected to narrow down the pool of viable candidate models describing UCOs, and ultimately provide decisive evidence regarding the true physical nature of the observed astrophysical black hole candidates \cite{lac:21,ccy:22}.

\acknowledgments
I would like to thank Robert Mann, Yasha Neiman, Jos{\'e} Senovilla, and Daniel Terno for useful discussions and helpful comments. This work was supported by the Quantum Gravity Unit of the Okinawa Institute of Science and Technology (OIST), an International Macquarie University Research Excellence Scholarship (IMQRES), and a Sydney Quantum Academy (SQA) Scholarship.

\end{document}